\documentclass[graybox]{svmult}

\usepackage{type1cm}        
\usepackage{makeidx}         
\usepackage{graphicx}        
\usepackage{multicol}        
\usepackage[bottom]{footmisc}

\usepackage{newtxtext}       %
\usepackage[varvw]{newtxmath}       

\makeindex             
\usepackage{float}

\begin{document}

\title*{Primordial Black holes: \\
  The asteroid mass window}
\author{Peter Tinyakov}
\institute{Peter Tinyakov \at Universit\'e Libre de Bruxelles,
  CP225 bvd. du Triomphe, 1050 Brussels, Belgium. \email{petr.tiniakov@ulb.be}
}
%
%
\maketitle

\abstract*{Primordial black holes (PBHs) are an attractive dark matter
  candidate, particularly if they can explain the totality of it. At PBH
  masses below $\sim 10^{17}$~g and above $\sim 10^{23}$~g this possibility is
  excluded from the variety of arguments and with different confidence. The
  range in between, often referred to as the ``asteroid mass window'',
  currently remains unconstrained. The most promising, in our view, way to
  probe this mass range is to use stars as the PBH detectors. If a star
  captures even a single PBH it starts being accreted onto it and eventually
  gets destroyed --- converted into a sub-solar mass black hole. This process
  may have a variety of signatures form a mere star disappearance to
  supernova-type explosions of a new kind. The viability of this approach
  depends crucially on the probability of PBH capture by stars. In this
  chapter we summarize the existing capture mechanisms and discuss their
  implications for constraining the abundance of (or perhaps discovering) PBHs
  in the asteroid mass window. }

\abstract{Primordial black holes (PBHs) are an attractive dark matter
  candidate, particularly if they can explain the totality of it. At PBH
  masses below $\sim 10^{17}$~g and above $\sim 10^{23}$~g this possibility is
  excluded from the variety of arguments and with different confidence. The
  range in between, often referred to as the ``asteroid mass window'',
  currently remains unconstrained. The most promising, in our view, way to
  probe this mass range is to use stars as the PBH detectors. If a star
  captures even a single PBH it starts being accreted onto it and eventually
  gets destroyed --- converted into a sub-solar mass black hole. This process
  may have a variety of signatures form a mere star disappearance to
  supernova-type explosions of a new kind. The viability of this approach
  depends crucially on the probability of PBH capture by stars. In this
  chapter we summarize the existing capture mechanisms and discuss their
  implications for constraining the abundance of (or perhaps discovering) PBHs
  in the asteroid mass window. }

\section{Introduction}
\label{sec:Intro}

Of the whole mass range where the primordial black holes (PBHs) may
potentially constitute the dark matter, the window of roughly
$10^{17}-10^{23}$~g is particularly difficult to assess. No actual
constraints, only proposals exist in this window at present
\cite{Carr:2020gox}.  Below $\sim 10^{17}$~g the PBHs evaporate fast enough to
either completely disappear by now or overproduce the diffuse gamma ray
background which is constrained by the existing observations
\cite{Carr:2009jm,Coogan:2020tuf}. At higher masses starting at around $\sim
10^{23}$~g (for comparison, the Moon mass is $7\times 10^{26}$~g) the PBHs
start being detectable through the gravitational lensing of background stars
\cite{PhysRevLett.111.181302,Griest:2013aaa}. The range between these two
values, spanning some 6 orders of magnitude and often referred to as the {\em
  ``asteroid mass window''}, is currently unconstrained.  In this mass range
the PBHs have sizes of order or smaller than the atomic size and are very
difficult to detect. 

Below $\sim 10^{23}$~g the PBHs are too light to be detectable by the standard
star lensing. A proposal has been made to use instead gamma-ray bursts as
lensed sources \cite{Barnacka:2012bm} and look, instead of time variability,
for features in the energy spectrum where PBHs, if present, would induce
oscillations. However, it has been argued later that finite size of
sources together with the wave optics effects will cause the decoherence
resulting in no detectable spectral modulations for most of the gamma-ray
bursts except, perhaps, a small fraction of rapidly varying ones
\cite{Katz:2018zrn}. Most of other proposals in the asteroid mass range turn
around using stars as PBH detectors. They are based either on the capture of
PBHs by stars with the subsequent star destruction
\cite{Capela:2013yf,Capela:2012jz,Capela:2014ita,Esser:2022owk,Esser:2023yut},
or on ignition of nuclear reactions in white dwarfs by a traversing PBH
\cite{Graham:2015apa}. An overview of these proposals can be found in
Ref.~\cite{Montero-Camacho:2019jte}. Here we focus on the PBH capture by
stars.

As we will see below, even in the most favorable conditions a star can only
capture a few PBHs over its lifetime. This is too small to have any dynamical
or gravitational effects on the star. However, once captured inside the star,
a PBH starts to accrete the ambient matter. If the Bondi accretion rate is
assumed\footnote{While not completely obvious, it is plausible that Bondi
  regime applies in the case of a small PBH inside the star. Nevertheless,
  this question requires a quantitative analysis which will be given
  elsewhere.}, the accretion is sufficiently rapid even in the case of
ordinary stars to be able to consume the whole star in a cosmologically short
time.  In the Bondi regime \cite{Bondi:1952ni}, the accretion rate by a PBH of
mass $m$ in the center of a star with the matter density $\rho_*$ is estimated
as
\[
\dot m \simeq 4\pi {G^2 m^2 \rho_* \over c_s^3} 
\]
where $G$ is the gravitational constant and $c_s$ the sound speed, and we have
suppressed a numerical factor of order 1 depending on the matter equation of
state. With this rate the accretion time is determined by the initial PBH mass
and reads
\[
t_\text{acc} = {c_s^3\over 4\pi G^2 \rho_*  m} \simeq
5\times 10^6 \text{yr} \left({c_s\over 500\text{km/s}}\right)^3
\left({150\text{g/cm}^3 \over \rho_*} \right) \left({ 10^{20}\text{g} 
\over m} \right). 
\]
We have substituted here the parameters typical of solar mass stars which have
lifetimes exceeding $\sim 10$~Gyr. We see that for PBH masses $m\gtrsim
10^{18}$~g the accretion times are much smaller than the age of the Universe,
and much smaller than the star lifetime. For compact stars such as white
dwarfs and neutron stars the accretion times are much shorter. 

Thus, capturing an asteroid-mass BH is fatal for stars --- they get destroyed
in a relatively short time. This may lead to several observable consequences
like disappearance of stars, creation of (sub)solar mass black holes and
supernova-type explosions of a new type. All three signatures may potentially
be used to constrain (or discover) the abundance of primordial black
holes. However, before discussing the signatures we need to consider how at
all PBHs can be captured inside a star.

\section{Capture --  generalities}
\label{sec:capture}

Being a part (or whole) of the dark matter, PBHs and stars are both
concentrated in galactic halos. However, because the interactions of a PBH
with matter are very weak relative to its enormous mass, it is very difficult
to settle a PBH inside a star. There are basically two ways: either a PBH can
be captured by a star from the local DM density, or it can get inside the star
from the star birth. In either case, the process typically requires two
stages:
\begin{itemize}
\item[(i)] ~{\em trapping}, or getting on a star-crossing gravitationally
bound orbit, and 
\item[(ii)] ~{\em cooling}, i.e., gradually loosing energy until settled
  completely inside the star.
\end{itemize}
The second stage, cooling, is common for both mechanisms. Once the PBH is
completely inside the star the accretion phase starts.

During the cooling stage, and also during the first stage in case of the
direct capture, the key quantity is the energy loss at a single star
crossing. So we start this section by having a closer, more quantitative
look at this quantity.

Once a PBH is on a bound star-crossing orbit it starts to lose energy at each
subsequent passage through the star. This process has to be fast enough, at
least it should not take longer than the star lifetime to get completely
inside the star. Given the PBH mass, the only free parameter here is the
initial size of the PBH orbit, depending on which there may or may not be
enough time to cool. Thus, not all trapped PBHs will eventually get captured
and destroy the star. The cooling process depends only on the energy loss; it 
will be the next topic in this section.

In the ideal case of an isolated star, once a PBH gets on a bound
star-crossing orbit it will cross the star again and again until it gets
absorbed inside. This is not necessarily the case for a star in a realistic
environment like a galaxy, because gravitational perturbations due to
neighboring stars may deviate the PBH and make it miss the star it is
orbiting, so that the energy loss will halt. So, as the last general point in
this section we will quantify the conditions under which the perturbers become
important.

\subsection{Energy loss}
\label{sec:enargy-loss}

A PBH passing through a star slows down for two reasons. First, it accretes
the star matter. In the star frame the total PBH momentum does not change,
while the mass grows, so the velocity decreases. This is equivalent to the drag
force 
\[
F_\text{acc} = - \dot m v
\]
where $\dot m$ is the mass accretion rate and we have assumed a
non-relativistic PBH velocity $v$. To estimate the mass accretion rate we note
that even those PBHs that have small (inbound) asymptotic velocities get
accelerated when they fall into a star, so that they cross it at a
(marginally) supersonic speed. Indeed, for a sun-like star, the escape
velocity from the star center is $v \sim \sqrt{3GM_*/R_*} \sim 750$~km/s,
while the sound speed in the Sun center is $\sim 500$~km/s. As one moves away
form the center towards the surface the sound speed decreases by an order of
magnitude, while the escape velocity changes by $\sqrt{2/3}$ at most.  For the
purpose of the estimate we may therefore consider the stellar matter as a
collection of free particles. Then the accretion rate is  $\dot m = \rho_*
\sigma_\text{BH} v$ where $\sigma_\text{BH}$ is the cross section of particle
capture by a black hole,
\[
\sigma_\text{BH} = 16\pi {G^2m^2\over v^2}. 
\]
We have therefore 
\[
F_\text{acc} = -16\pi \rho_* G^2m^2  
\]
where the velocity $v$ canceled out. 

The second force that slows down the PBH is the dynamical friction
\cite{RevModPhys.21.383,1987gady.book.....B}. This force arises because matter
particles get dragged along with the passing PBH by the gravitational force;
correspondingly, the PBH is slowed down. The dynamical friction force is
written as follows,
\begin{equation}
F_\text{dyn} = - 4\pi \rho_* G^2 m^2 {\ln\Lambda \over v^2}
\label{eq:Fdyn}
\end{equation}
where $\ln\Lambda \sim 30$ is a Coulomb logarithm. 
Thus, the dynamical friction is dominant over the accretion by the factor 
$\ln\Lambda/ v^2$ which is much larger than 1, except perhaps
in the case of neutron stars where $v\sim 1$ and the dynamical friction is
somewhat reduced due to matter degeneracy. 

As in the case of accretion, eq.~(\ref{eq:Fdyn}) assumes that the matter
particles do not interact among themselves, which is a good approximation at
supersonic PBH velocities. In our case the velocity of the PBH when passing
through a star is only marginally supersonic. In this case the dynamical
friction (\ref{eq:Fdyn}) gives a smaller force than a more accurate treatment
that takes into account properties of the medium
\cite{1999ApJ...513..252O}. The accurate answer is some factor $\sim 2$ larger
than eq.~(\ref{eq:Fdyn}) in the case of Sun-like stars (see Fig.4 of
Ref.~\cite{Oncins:2022djq}) due to a resonance-type enhancement at velocities
close to the sound speed \cite{1999ApJ...513..252O}. We do not include this
numerical factor in what follows.

With the force (\ref{eq:Fdyn}), the energy loss per single star crossing is
$\Delta E \sim R_* F_\text{dyn}$. More accurately, one has to average the
energy loss over the PBH impact parameter. If one assumes for simplicity the
uniform distribution of the PBH trajectories in the plane perpendicular to the
PBH velocity we get, in agreement with eq.(2) of Ref.\cite{Capela:2013yf}
\begin{equation}
E_\text{loss} = {1\over \pi R_*^2} \int ds\, dl F_\text{dyn} =
{4 M_* G^2 m^2  \over R_*^2 v^2}\ln\Lambda. 
\label{eq:Eloss-1}
\end{equation}
Here we have ignored the change of $v^2$ along the PBH trajectory within
the star. We will see later, however, that the distribution of trajectories in
some cases is not uniform, and eq.~(\ref{eq:Eloss}) in these cases is only an
order-of-magnitude estimate. 

Substituting in (\ref{eq:Eloss-1}) the escape velocity from the star surface 
$v^2_\text{esc} = 2GM_*/R_*$
we can rewrite it as follows,
\begin{equation}
{E_\text{loss}\over m} = {2Gm\over R_*} \ln\Lambda = 
{R_g\over R_*} {m \over M_*} \ln\Lambda ,
\label{eq:Eloss}
\end{equation}
where $R_g = 2GM_*$ is the gravitational radius of the star. Note that this expression
is only valid in the case when $v \simeq v_\text{esc}$, which explains why the 
independence of $E_\text{loss}$ of $M_*$ is not a problem. We see from
eq.~(\ref{eq:Eloss}) that the more compact is the star, the larger are the
energy losses, but even in the best case of a neutron star with $R_g\sim
R_*/3$ the energy loss is very small for an asteroid mass PBH for which $m/M_*
\lesssim 10^{-10}$.

It follows immediately from this estimate that only very slow PBHs can be
trapped after one star crossing. To be trapped, the asymptotic kinetic energy
of a PBH must be smaller than $E_\text{loss}$, which translates into the condition
on the asymptotic velocity $v_\infty$ of PBH 
\[
v_\infty < 2 \left( {Gm\over R_*}\ln\Lambda\right)^{1/2}.
\]
Numerically, a PBH of mass $m=10^{23}$~g 
can only get trapped if its asymptotic velocity relative to the star does not
exceed some maximum value which is $0.03$~km/s for the Sun and $9$~km/s for a
neutron star. Even in the case of a neutron star this velocity is smaller
than the velocity dispersion in most astrophysical environments, with notable
exception of some dwarf galaxies.

Another immediate consequence of this estimate is large initial size of the
PBH bound orbit after trapping. When a PBH from the ambient DM distribution
passes through a star it loses the amount of energy (\ref{eq:Eloss}). If its
initial kinetic energy was smaller it gets trapped. A typical energy of a
freshly trapped PBH is therefore of the order $-E_\text{loss}$. Such PBHs have
radial orbits of the size
\begin{equation}
r_\text{max} \sim {GM_* m \over E_\text{loss}} 
= {M_*\over 2 m \ln\Lambda} R_*.
\label{eq:r_max}
\end{equation}
The trapped PBH continues to lose energy at each subsequent star crossing and
finally gets completely absorbed in the star, provided that (i) there is
enough time and (ii) the trajectory is not perturbed so that it stops passing
through the star and lose energy. We come now to these two conditions.

\subsection{Cooling time}
\label{sec:cooling-time}

Since the energy loss at a single star crossing is very small, the size of the
first bound orbit is very large, cf. eq.~(\ref{eq:r_max}). Outside of the star
such an orbit is a Keplerian orbit with the apastron $a =r_\text{max}$ and
periastron $b< R_* \ll a$. One may therefore consider this orbit as radial. For
such an orbit the time between the first and the next star crossing is
\[
\Delta t = \pi {a^{3/2}\over \sqrt{R_g}}.  
\] 
If we assume that after the first passage the PBH total energy is
$E_1= -E_\text{loss}$, at the $n$-th step its energy is 
\[
E_n = -n E_\text{loss}.
\]
Correspondingly, the size of the $n$-th orbit is 
\[
a_n = {r_\text{max} \over n} 
\]
and the return time at the $n$-th step is 
\[
\Delta t_n = {\Delta t \over n^{3/2}}. 
\]
The total cooling time converges and is
saturated by a few first orbits. Summing up $\Delta t_n$  one finds 
\begin{equation}
t_\text{cool} = \sum_{n=1} \Delta t_n =  \pi \zeta(3/2) {r_\text{max}^{3/2}\over
  \sqrt{R_g}} \sim 8 {r_\text{max}^{3/2}\over
  \sqrt{R_g}}.
\label{eq:t_cool}
\end{equation}
This equation is valid when the initial size of the orbit is $r_\text{max}$
given by (\ref{eq:r_max}), as determined by $E_\text{loss}$ of
eq.~(\ref{eq:Eloss}).

In the case when the initial size of the orbit $a_0$ is much smaller than
$r_\text{max}$, one may  either calculate a partial sum in
eq.~(\ref{eq:t_cool}) and then use the asymptotics of the resulting Hurwitz
zeta function, or use the continuous energy loss approximation. After the star
crossing, the energy changes by $-E_\text{loss}$, and correspondingly the
apastron of the orbit decreases by
\[
\Delta a = - {a^2 \over GM_* m} E_\text{loss}. 
\]
Approximating $ da/dt$ by $\Delta a/\Delta t$ one obtains the differential
equation for the change of the apastron
\[
{da\over dt} = - {2 E_\text{loss}\over \pi m\sqrt{R_g} } \sqrt{a}. 
\]
Solving this equation one finds the cooling time as a function of the initial
apastron $a_0$,
\begin{equation}
t_\text{cool}(a_0) = {\pi \sqrt{R_g} m \over E_\text{loss}} \sqrt{a_0}
= {\pi M_* R_* \over m \sqrt{R_g} \ln\Lambda} \sqrt{a_0}, 
\label{eq:t_cool-small_a}
\end{equation}
where we have substituted $E_\text{loss}$ from eq.~(\ref{eq:Eloss}).  This
equation is valid for $a\ll r_\text{max}$, with $r_\text{max}$ given by
eq.~(\ref{eq:r_max}). Note that if one formally takes $a_0 = r_\text{max}$ in
eq.~(\ref{eq:t_cool-small_a}), the result is parametrically the same as
eq.~(\ref{eq:t_cool}) and only differs by a numerical coefficient.

\subsection{Perturbers}
\label{sec:perturbers}

For a successful cooling the bound PBH trajectory must cross the star at each
approach, i.e. the periastron must be smaller than the star radius. For an
isolated star and PBH orbits that become bound as a result of the energy loss
{\em inside the star} this is automatically guaranteed. However, in a realistic
environment (in a galaxy) the gravitational perturbations from nearby stars
may make the trajectory that originally passed through the star to miss it at
the next approach. If this happens, the PBH starts orbiting the star without
energy loss and the cooling will never complete. Following
Ref.~\cite{Esser:2022owk} and assuming the perturber is a nearby star of the
same mass, let us derive the condition on the distance to this star which
guarantees that the perturbation is not big enough to halt the cooling.  

For very extended nearly radial orbit the periastron $r_{\rm min}$ is
completely determined by the PBH angular momentum $J$ 
through the relation
\[
r_{\rm min} = {J^2/ (m^2 R_g)}.
\]
Requiring $r_{\rm min}<R_*$ sets an upper bound on the angular momentum
\begin{equation}
J_{\rm max}/m = \sqrt{R_* R_g}.
\label{eq:J_max}
\end{equation}
The condition that the orbit crosses the star $r_{\rm min}<R_*$ is thus
equivalent to requiring that $J<J_\text{max}$.

Consider the change of the PBH angular momentum under the influence of a small
gravitational potential $U(\vec x)$. For definiteness we suppose that the
unperturbed PBH trajectory is $x(t)$, and the perturbation is in the plane
$(x,y)$. We are interested in the change of the angular momentum $\Delta J$
accumulated over one fall from $x=r_\text{max}$ to $x\sim R_*$. We have
\[
\Delta J / m = \int _0^{T/4} x^2(t)
       {\partial^2 U\over \partial x \partial y} dt,
\]
where $T$ is the period of the radial Keplerian orbit with the apastron
$r_{\rm max}$. To estimate the integral we use the unperturbed motion $x(t)$
and assume that the perturber is far away, so that the derivative $\partial_x
\partial_y U$ is approximately constant along the trajectory. This gives
\begin{equation}
  \Delta J / m = {5\pi\over 16}{r_{\rm max}^{7/2}\over R_g^{1/2}}
         {\partial^2 U\over \partial x \partial y}. 
\label{eq:DeltaJ}
\end{equation}
We have to require that $ \Delta J \lesssim J_\text{max}$. 

If the perturber is a star with the same mass at a distance $d\gg r_{\rm
  max}$ which we assume to be static, this condition translates into
\begin{equation}
r_{\rm max} < d \left({R_*\over \alpha d}  \right)^{1/7}. 
\label{eq:constraint2}
\end{equation}
Here $\alpha$ is a numerical coefficient of order 1 that depends on the
relative orientation of the PBH orbit and the direction to the perturber. In
view of the small power in eq.~(\ref{eq:constraint2}) the
numerical value of this coefficient is not so important. In a given
environment the distance $d$ can be estimated from the density of stars.

\section{Direct capture}
\label{sec:direct-capture}

With these tools at hand, consider now the direct capture in more detail. It
is instructive to start by deriving the rate of trapping of PBHs by an
isolated star. So, assume an isolated star at rest is placed in a DM halo of
density $\rho_\text{DM}$ consisting of PBHs which have Maxwell distribution in
velocities with the dispersion $\sigma$,
\begin{equation}
  dn = {\rho_\text{DM} \over m}
  \left({3\over 2\pi \sigma^2} \right)^{3/2}
  \exp\left( - {3v^2 \over 2\sigma^2 } \right) d^3v. 
  \label{eq:Maxwell}
\end{equation}
Some of the PBHs from the halo will cross the star and lose enough energy to
become gravitationally bound on orbits that cross the star, i.e., will
get trapped. Averaging over
the Maxwell distribution, the rate at which this happens is
\cite{Press:1985ug}
\[
  { dN\over dt} = 2\sqrt{6\pi} {\rho_\text{DM} \over \sigma m} GM_*R_*
  \left\{1-\exp \left(- {3E_\text{loss}\over m \sigma^2} \right)
  \right\} \simeq 6 \sqrt{6\pi}{\rho_\text{DM} \over \sigma^3}
  {GM_*R_*\over m^2}E_\text{loss}, 
\]
where in the last equality  we have expanded the exponential 
assuming $3 E_\text{loss}/(m \sigma^2)\ll 1$. Note that only the low-velocity
end of the distribution contributes at small $E_\text{loss}$, so the details
of the velocity distribution in fact do not matter.

If we substitute here the expression (\ref{eq:Eloss}) for 
$E_\text{loss}$ we find
\[
{ dN\over dt} = 24 \sqrt{6\pi} {\rho_\text{DM} \over \sigma^3}
G^2 M_* \ln\Lambda. 
\]
Remarkably, when applicable, the result does not depend neither on the PBH
mass nor on the star radius. Numerically, the trapping rate reads
\begin{equation}
{ dN\over dt} = 0.5~ \text{Gyr}^{-1}
\left( {7\text{km/s}\over \sigma }\right)^3
\left({\rho_\text{DM} \over 100~\text{GeV/cm}^3}\right)
\label{eq:dN/dt-fin}
\end{equation}
for $M_*=M_\odot$.  We stress again that this is the rate at which the PBHs
get trapped (settle on bound star-crossing orbits) after the first collision
with the star. While only these PBHs may repeatedly collide with the star
and eventually lose enough energy to completely sink inside, not all
of them will necessarily do it. Therefore, this is the upper bound; the actual
capture rate can only be smaller. This expression has been derived from
different arguments in Ref.\cite{Montero-Camacho:2019jte}\footnote{There is a
  factor 2 difference between eq.~(\ref{eq:dN/dt-fin}) and
  Ref.~\cite{Montero-Camacho:2019jte} which we attribute to a different
  estimate of $E_\text{loss}$.}

This bound implies that by the direct mechanism it is possible to capture a
few PBHs at most during the whole lifetime of the Universe. It applies to all
types of stars except perhaps neutron stars for which the approximations which
we have used become marginal, so a numerical factor may arise. We will
consider the case of neutron stars in more detail below.

\subsection{Ordinary stars}
\label{sec:direct-capt-ordin}

Let us show that ordinary stars are very inefficient in capturing PBHs
directly. In this case the actual capture rate is much smaller than 
given by eq.~(\ref{eq:dN/dt-fin}) because of further suppression resulting
from cooling time and perturbation constraints.

Consider first the cooling time (\ref{eq:t_cool}). As explained above, it is
determined by the initial orbit size $r_\text{max}$ which is in turn
determined by the energy loss. Since the initial total energy of a PBH is
positive, the smallest (most negative) value the energy may have after the
star crossing is $-E_\text{loss}$. The larger $E_\text{loss}$, the smaller is
$r_\text{max}$ and the shorter the cooling time.

The energy loss varies depending on the impact parameter and is maximum for
the trajectory passing through the star center. But even for such a trajectory
the energy loss is ony factor $3/2$ larger than the average value
(\ref{eq:Eloss}).  With the average energy loss, the initial orbit size is
\[
r_\text{max} = 3.7~\text{kpc} \left({10^{20}~\text{g}\over m} \right),
\]
and the corresponding cooling time is 
\[
t_\text{cool} = 2\cdot 10^{13}~\text{yr} 
\left({10^{20}~\text{g}\over m} \right)^{3/2}.
\]
We see that the initial size of the orbit is unrealistically large except
maybe at the upper end of the asteroid window $m\sim 10^{23}$g. The cooling
time exceeds $10$~Gyr in the whole mass window, beign comparable to it only at
the upper end. 

The constraint from perturbers cannot be satisfied either. As we will discuss
below in Sect.~\ref{sec:expected-constraints}, the most promising known
systems maximizing the capture rate (\ref{eq:dN/dt-fin}) are ultra-faint dwarf
galaxies where the star-to-star distance $d$ is of the order $d\sim 6
\text{pc}$. The maximum initial orbit size has to be at least an order of
magnitude smaller: making use of eq.~(\ref{eq:r_max}) with $\alpha=1$ we find
\[
r_\text{max} < 0.4~\text{pc}. 
\]
Again, this condition cannot be satisfied unless $m > 10^{24}$~g
(cf. eq.~(\ref{eq:r_max})), which is outside of the mass window we consider.

We conclude that direct capture does not work for ordinary Sun-like
stars. Even if a fraction of PBHs escapes gravitational perturbations by
chance, there is not enough time to lose the energy. 

\subsection{Neutron stars}
\label{sec:direct-capt-neutr}

In the case of neutron stars there are several differences which have to be
taken into account. First, a neutron star has radius only about factor $\sim
3$ larger than its Schwarzschild radius. This means the General Relativity
(GR) effects start becoming non-negligible. They lead, in particular, to a
marginal enhancement by a factor $1/(1-R_g/R_\text{NS})\approx 1.6$ in the
capture rate \cite{Kouvaris:2007ay}.

A PBH falling onto a neutron star accelerates to a semi-relativistic velocity
$v\sim \sqrt{R_g/R_\text{NS}} \sim 0.6$. This velocity is similar or larger
than the sound speed in the neutron star core which is most often cited within
the $0.3-0.6$ range, but is not well known and depends on the neutron star
model.  For semi-relativistic velocities the dynamical friction
(\ref{eq:Fdyn}) loses its $1/v^2$ enhancement.

Another complication is that nuclear matter is degenerate, and therefore in
the approximation of non-interacting particles (neutrons) not all scattering
states can be excited, but only those for which the momentum transfer exceeds
the Fermi momentum. Since the dynamical friction force can be viewed as a 
result of the momentum transfer to scattered matter particles, less
scattering means weaker friction force. 

The calculation of the dynamical friction that takes into account both effects
can be found in the Appendix of Ref.~\cite{Capela:2013yf} where the
contribution of the force due to the direct accretion was also included. The
resulting energy loss, averaged over different impact parameters of PBH
trajectories passing through the star, was found to be 
\begin{equation}
E_\text{loss} = {4 G^2 m^2 M_*\over R_*^2} 
\left\langle {\ln\Lambda\over v^2} \right\rangle 
\label{eq:E_loss_NS}
\end{equation}
where the $\langle ...\rangle$ denotes the density-weighted average 
over the star volume. Numerically one finds, using a realistic neutron star
density profile, 
\[
\left\langle {\ln\Lambda\over v^2} \right\rangle \simeq 15. 
\] 
Remarkably, accounting for the degeneracy only reduces the energy loss by a
factor $\sim 6$ as compared to the naive estimate $\ln\Lambda/v^2 \sim 83$. 

Apart from the dynamical friction, one should include in the energy loss the
tidal effects, excitation of the surface waves and gravity wave
emission. Numerically these effects turn out not very important for PBH masses
in the asteroid window. With the account of all the known energy losses, the
direct capture rate by {\em an isolated} neutron star was calculated in
Ref.~\cite{Genolini:2020ejw}. The result can be written as follows,
\begin{equation}
{dN\over dt} =  0.26\, \text{Gyr}^{-1} 
\left( {7\text{km/s}\over \sigma }\right)^3
\left({\rho_\text{DM} \over 100~\text{GeV/cm}^3}\right) {\cal C}(X),
\label{eq:direct-cap-NS}
\end{equation}
where we have included the GR enhancement by a factor
$\approx 1.6$ \cite{Kouvaris:2007ay}. 
Here the variable $X$ is
\[
X = 0.018 \left({m\over 10^{20}\text{g}}\right)
\left( {7\text{km/s}\over \sigma }\right)^2. 
\]
and the function ${\cal C}(X) \approx 1 $ in the range $X\in [0,20]$ and drops
as power law for larger values of $X$. Note
that $X<20$ for PBH masses up to $10^{23}$~g at $\sigma =
7$~km/s. The full shape of ${\cal C}(X)$ can be found in
Ref.~\cite{Genolini:2020ejw}, Fig.~3. 

Let us now check the conditions coming from the cooling time and the
gravitational perturbations by nearby stars. For both conditions the key
quantity is the size of the initial bound orbit $r_\text{max}$. Given the
energy loss by a PBH passing through the neutron star (\ref{eq:E_loss_NS}),
the initial size of the orbit can be found from eq.~(\ref{eq:r_max}) which
gives
\begin{equation}
r_\text{max} = {R_*^2 \over 4 G m}
\left\langle {\ln\Lambda\over v^2} \right\rangle^{-1}
= 0.7\,\text{pc} \left({ 10^{20} \text{g}\over m}\right).
\label{eq:r_max_NS}
\end{equation}
In terms of $r_\text{max}$, the cooling time is given by eq.~(\ref{eq:t_cool})
which contains no other parameters. This gives
\[
t_\text{cool} = 5\cdot 10^7\,\text{yr} \left({ 10^{20} \text{g}\over
  m}\right)^{3/2}.
\]
We see that the cooling is sufficiently fast as long as $m \gtrsim 3\cdot
10^{18}\,\text{g}$, i.e. almost in the whole asteroid mass range except the
lightest masses.

The condition that nearby stars do not spoil the cooling is given by
eq.~(\ref{eq:constraint2}). We may use the same estimate of the star-to-star
distance as above, $d\sim 6$~pc, as found in some of the prominent dwarf
galaxies. Making use of this value and $R_*\simeq 10$~km for the neutron star
radius we find that the condition (\ref{eq:constraint2}) requires
\[
r_\text{max} < 0.08\,\text{pc}.
\]
Combining this with the expression for $r_\text{max}$ as a function of $m$
eq.~(\ref{eq:r_max_NS}), this condition translates to
\begin{equation}
m \gtrsim 10^{21}\,\text{g}. 
\label{eq:perturber_NS}
\end{equation}
Thus, the  high mass end of the asteroid mass window satisfies this condition,
while the low mass end does not. 

In summary, both conditions together are satisfied at the high mass end of the
asteroid window, eq.~(\ref{eq:perturber_NS}). For these masses the capture
rate is given by eq.~(\ref{eq:direct-cap-NS}). The rate is not vanishingly
small only in environments with high DM density and low velocity dispersion, 
similar to dwarf galaxies where, numerically, the capture rate may reach
values of 1 per a few Gyr.

\section{Capture at birth}
\label{sec:capture-at-birth}

Let us now turn to the second mechanism --- capture of PBHs by stars at their
birth. Stars form from dense gas clouds which at some point lose their
stability with respect to gravitational collapse. At the initial stages these
clouds are almost uniform. As the collapse progresses a dense protostellar
object develops in the center \cite{1996ApJ...469..366B,Bate:1998hg} which
starts accreting the surrounding gas, forming at the end a
pre-main-sequence star. This process happens at the Kelvin-Helmholtz timescale
which is much longer than the free fall time. 

Star formation takes place in galaxies in the presence of DM density
$\rho_\text{DM}$ which we will assume, as above, to consist of PBHs. We will
also assume that the PBHs have Maxwell distribution in velocities with the
velocity dispersion $\sigma$. A typical baryonic density in a cloud
of $10^6-10^7$~GeV/cm$^3$ is much larger than the DM density in the galactic
halos, so whatever happens to PBHs has no influence on the cloud dynamics.

The capture mechanism is the following. A gas cloud from which a star will
form is gravitationally bound and also has some amount of PBHs gravitationally
bound to it. When the cloud collapses into a pre-main-sequence star, the
time-dependent gravitational potential of the collapsing baryons drags these
PBHs along, so that their distribution becomes peaked at the star
position. When the star is finally formed, there is a probability to find a
PBH on a bound orbit either completely contained inside or passing through the
star. The fist ones are already captured; the second start cooling and will be
eventually captured if the cooling is successful.

Obviously, only the trajectories of PBHs that spend large time inside the
contracting cloud are significantly affected by the contraction, so the PBHs
that are not bound to the cloud can be neglected. The number of captured PBHs
is proportional to their total number in the corresponding part of the phase
space. This number in turn is proportional to the total DM density
$\rho_\text{DM}$, and to the volume of the velocity space limited from above
by the escape velocity from the cloud.  The escape velocity can be estimated
from the baryonic density $\rho_0$ and radius $R_0$ of the cloud. These
parameters are known from observations \cite{Kirk:2005ng} and are summarized
in Ref.\cite{Capela:2012jz}. For example, for a star of $1\,M_\odot$ the
values are $R_0=4300$~AU and $\rho_0=10^6$~GeV/cm$^3$. For these parameters
the escape velocity is $v_\text{esc} = 0.6$~km/s. This is much smaller than
the velocity dispersion in known dwarf galaxies. Therefore, the volume in the
DM velocity space corresponding to trajectories bound to the cloud is
proportional to $v^3_\text{esc}/\sigma^3$. We conclude that the number of
bound trajectories, and therefore the number of captured PBHs is proportional
to the combination ${\rho_\text{DM}/ \sigma^3}$, similarly to the direct
capture case.

\subsection{Mean captured number: ordinary stars}
\label{sec:mean-N-sun}

Consider first ordinary stars because in any case compact stars --- white
dwarfs and neutron stars --- are formed from main sequence stars and therefore
pass first through this stage. So, the PBHs captured by compact stars at birth
are those that they inherit from their main sequence progenitors. 

In order to calculate the total number of captured PBHs one has to know the
distribution of the PBHs in position and velocity after the star is formed.
This distribution builds up from bound PBH trajectories during the star
formation.  The fact that the formation happens slowly on a time scale much
longer than the free fall time implies that the motion of the bound PBHs can
be described in the adiabatic approximation. This means that the details of
the contraction process do not matter, only the initial and final state.

The {\em maximum} number of PBHs that can be captured by a star at birth
can be estimated by assuming that after the star is formed the phase space of
PBHs is filled with the maximum density all the way up to positive energies.
From the Liouville theorem, the phase space density after the star formation
can nowhere exceed the maximum phase space density before --- that is, the one
at low velocities, assuming the Maxwellian DM distribution. From
eq.~(\ref{eq:Maxwell}), this phase space density is
\begin{equation}
{\cal Q} = {\rho_\text{DM} \over m}
  \left({3\over 2\pi \sigma^2} \right)^{3/2}. 
\label{eq:Q-density}
\end{equation}
We are not, however, aware of the proof that this phase space {\em has} to be
filled. It is straightforward to argue that the Liouville theorem and the
angular momentum conservation together do not guarantee this. A 
counter-example is a very fast (instantaneous) contraction which satisfies both
properties but obviously does not populate the circular orbits close to the
star.

It is instructive nevertheless to make an analytic estimate (the detailed
calculation can be found in Ref.~\cite{Oncins:2022djq}). Starting
from the distribution that has constant phase space density
(\ref{eq:Q-density}), we want to calculate the number of orbits that (i) are
bound and (ii) have periastrons smaller than the star radius $R_*$. In other
words, we have to integrate
\begin{equation}
dN = {\cal Q} d^3r d^3v = 8 \pi^2 {\cal Q} r^2 v_\phi dr dv_r dv_\phi
\label{eq:dN-start}
\end{equation}
over the part of the phase space that satisfies (i) and (ii). Here we have used
symmetries to reduce the number of variables to three: radius $r$, radial
velocity $v_r$ and the modulus of the tangential velocity $v_\phi$. 

The shape of the Keplerian orbits is completely characterized by conserved
energy $E$ and angular momentum $L$,
\begin{eqnarray}
\label{eq:keplerE} 
\epsilon &\equiv & {E\over m} = {1\over 2} (v_r^2 + v_\phi^2)
- {GM_*\over r},\\
\label{eq:keplerL} 
l &\equiv & {L\over m} = rv_\phi. 
\end{eqnarray}
At fixed $r$ one may trade the velocities $v_r$ and $v_\phi$ for $\epsilon$
and $l$. From eqs.~(\ref{eq:keplerE})--(\ref{eq:keplerL}) the Jackobian of this
transformation is
\[
dv_r dv_\phi = {1\over r v_r} d\epsilon dl . 
\]
One has therefore from eq.~(\ref{eq:dN-start})
\[
dN = 8 \pi^2 {\cal Q} {l\over v_r}  dr d\epsilon dl .
\]
We may express the radial velocity through $\epsilon$ and $l$ as follows,
\[
v_r = \pm \sqrt{  {2GM_*\over r} + 2\epsilon - {l^2\over r^2} }
\]
where both signs are physical. We find therefore
\begin{equation}
dN = 8 \pi^2 {\cal Q} {1\over \sqrt{  {2GM_*\over r} + 2\epsilon - {l^2\over
      r^2} }}  dr d\epsilon d(l^2).
\label{eq:N-step1}
\end{equation}
The extra factor 2 in this equation is introduced to account for two signs of
$v_r$ at given $E$ and $L$. 

One more change of variables is needed to make the integration trivial. Since
the condition (ii) is imposed on the periastron, it is useful to change the
variables from $\epsilon$ and $l^2$ to the periastron $b$ and
apastron $a$. For any finite $a$ the condition (i) --- negative total energy
-- is then automatically satisfied. We will see at the end that most of the
contributions come from very elongated orbits for which $a\gg b$. In this case
one can approximate $a$ and $b$ as follows,
\begin{eqnarray}
\label{eq:kepler-a}
a &=& - {GM_*\over \epsilon} + {\cal O}(b/a),\\
\label{eq:kepler-b}
b &=& {l^2\over 2GM_*} + {\cal O}(b/a). 
\end{eqnarray}
Recall that the energy is negative, $\epsilon<0$. In the same approximation,
one can neglect the last term under the square root in
eq.~(\ref{eq:N-step1}). Eqs.~(\ref{eq:N-step1}), (\ref{eq:kepler-a}) and
(\ref{eq:kepler-b}) then give
\[
  dN = 4\pi^2 {\cal Q} { R_g^{3/2} r \over a^{3/2} \sqrt{ar-r^2} } dr\, da\,
  db, 
\]
where as before $R_g = 2GM_*$ is the Schwarzschild radius of the star. 

We will need at the end the number of PBHs on star-crossing orbits with the apastron
smaller than a given $a_\text{max}\gg R_*$. This makes the integral
convergent. We therefore have to integrate the
above expression over $r$ from $b$ to $a$, then over $a$ from $b$ to
$a_\text{max}$ and finally over $b$ from 0 to $R_*$,
\[
N = 4\pi^2 {\cal Q} R_g^{3/2} \int_0^{R_*} db \int_b^{a_\text{max}} { da \over
  a^{3/2} } 
\int_b^a {r \, dr \over  \sqrt{ar-r^2} }.
\]
One may notice that the dominant contribution to this integral comes from the
region $r,a\gg R_*$. We may therefore change the lower limit of the
integration over $r$ to 0 instead of $b$. The integral over $r$ then gives
\[
\int_0^a { r \over  \sqrt{ar-r^2} } dr = {a\pi\over 2}, 
\]
so that 
\begin{equation}
N= 2\pi^3 {\cal Q}  R_g^{3/2}
\int_0^{R_*} db \int_b^{a_\text{max}} 
{ da \over \sqrt{a} }
\label{eq:dN}
\end{equation}
The integral over $a$ now becomes explicitly convergent at the lower limit
and we can again replace $b$ by $0$. Then the integration gives
\begin{equation}
N = 4\pi^3 {\cal Q} R_g^{3/2} R_* \sqrt{a_\text{max}}= 
{\rho_\text{DM} \over m \sigma^3} (3\pi R_g)^{3/2} R_* \sqrt{2a_\text{max}}.
\label{eq:N-final}
\end{equation}
This expression agrees with the one obtained in Ref.\cite{Oncins:2022djq} for
the same quantity. It also agrees with the number of PBHs that settle on the
star crossing orbits calculated numerically in Ref.\cite{Esser:2022owk}. This
implicitly confirms our starting assumption that when the star is formed, the
whole phase space with $E<0$ is filled with the maximum phase-space density.

The captured number grows infinitely as $a_\text{max}\to\infty$, i.e. when
larger and larger orbits are included. At some point, however, $a_\text{max}$
becomes so big that either the cooling time becomes longer than the age of the
star, or the perturbations from nearby stars can no more be neglected. We will
see below that, in realistic conditions of dwarf galaxies, the cooling
requirement is more constraining at masses below $\sim 10^{21}$~g. In order to
estimate the number of captured PBHs from eq.~(\ref{eq:N-final}) in this mass
range it remains, therefore, to substitute in eq.~(\ref{eq:N-final}) the
maximum size of the orbit that still has time to cool and settle inside the
star. The orbits that PBHs acquire after the contraction are much smaller than
the critical value (\ref{eq:r_max}), so we should use
eq.~(\ref{eq:t_cool-small_a}) for the cooling time.  Comparing
eqs.~(\ref{eq:N-final}) and (\ref{eq:t_cool-small_a}) we see that their
dependence on $a$ and $m$ is the same, so that we can rewrite the captured
number $N$ as follows,
\begin{equation}
N = 3\sqrt{6\pi} R_g^2 {\ln\Lambda\over M_*}\, {\rho_\text{DM}\over \sigma^3} 
\,t_\text{cool}. 
\label{eq:N-tcool}
\end{equation}
From this expression we conclude that when the capture is limited by the
cooling time, the number of captured PBHs does not depend on the PBH mass $m$.
Note, however, that eq.~(\ref{eq:t_cool-small_a}) is approximate, so the
estimate (\ref{eq:N-tcool}) is only valid when the former is correct. 

If we substitute in eq.~(\ref{eq:N-tcool}) the cooling time of order of the
star lifetime, $ t_\text{cool} \sim 10$~Gyr, we find that numerically
\begin{equation}
N = 2.3 \, \left({\rho_\text{DM}\over 100\,\text{GeV/cm}^3}\right)
\left(7\,\text{km/s}\over \sigma\right)^3
\left({t_\text{cool}\over 10\,\text{Gyr}}\right).
\label{eq:Ncap-num}
\end{equation}
This is very close to the result of the numerical simulations of
Ref.\cite{Esser:2022owk} as will be discussed below. It should be stressed,
however, that several approximations have been made to obtain
eq.~(\ref{eq:N-tcool}), so the numerical agreement of eq.~(\ref{eq:N-tcool})
with simulations is a mere coincidence. Among these approximations we have
used the {\em average} energy loss and assumed, when calculating the average
in eq.~(\ref{eq:Eloss-1}), that the trajectories are distributed uniformly
over the impact parameter plane inside the star. We see now from
eq.~(\ref{eq:N-final}) that this is not the case: if it were, the measure in
eq.~(\ref{eq:dN}) would have been $b\,db$, not $db$. We have also assumed that
$b\ll a$.  This approximation may break for small PBH masses leading to small
energy loss, cf. eq.~(\ref{eq:Eloss-1}), which would imply that only
relatively compact trajectories have time to cool down.

If one does not want to rely on assumptions, numerical methods must be used to
determine $N$. Such simulations have been performed in
Refs.\cite{Capela:2012jz,Capela:2014ita,Esser:2022owk}. The calculation is
greatly simplified by the fact that the total dark matter mass on bound orbits
is by many orders of magnitude smaller than the star mass, so one can simulate
PBH trajectories {\em one by one} in the same gravitational potential of the
contracting cloud, each time with different initial conditions. Further
simplification is due to the adiabatic character of the contraction. Because
of that one may chose a convenient parameterization of the time-dependent
cloud shape --- all choices should give the same result as long as the
contraction is slow.
 
The general algorithm is as follows (see Ref.~\cite{Esser:2022owk} for
details). One generates random initial conditions for PBH trajectories
covering uniformly the relevant region of the phase space. For numerical
efficiency one may sample only the region of negative starting energy and
small angular momentum containing (almost) all the trajectories that, by the
end of the contraction, may become gravitationally bound and passing through
the star. Each trajectory is then evolved in the gravitational field of the
contracting cloud. The trajectories that do pass through the star at the end
of the contraction are selected.

Among these trajectories, the ones with the apastron  too large to survive
the deflections by nearby stars (see Sect.~\ref{sec:perturbers}) are then
discarded. For each of the remaining trajectories the cooling time has to be
checked. It is not sufficient at this stage to rely on the average energy loss
(\ref{eq:Eloss}).  In Ref.\cite{Esser:2022owk} the energy losses were
calculated with the account of the actual star density profile (the Solar
profile was used) and the parameters of individual trajectories. The
trajectories having the cooling time longer than the star lifetime are again
discarded. The remaining ones are considered successful.

\begin{figure}[ht]
  \centering
  \includegraphics[width=0.8\textwidth]{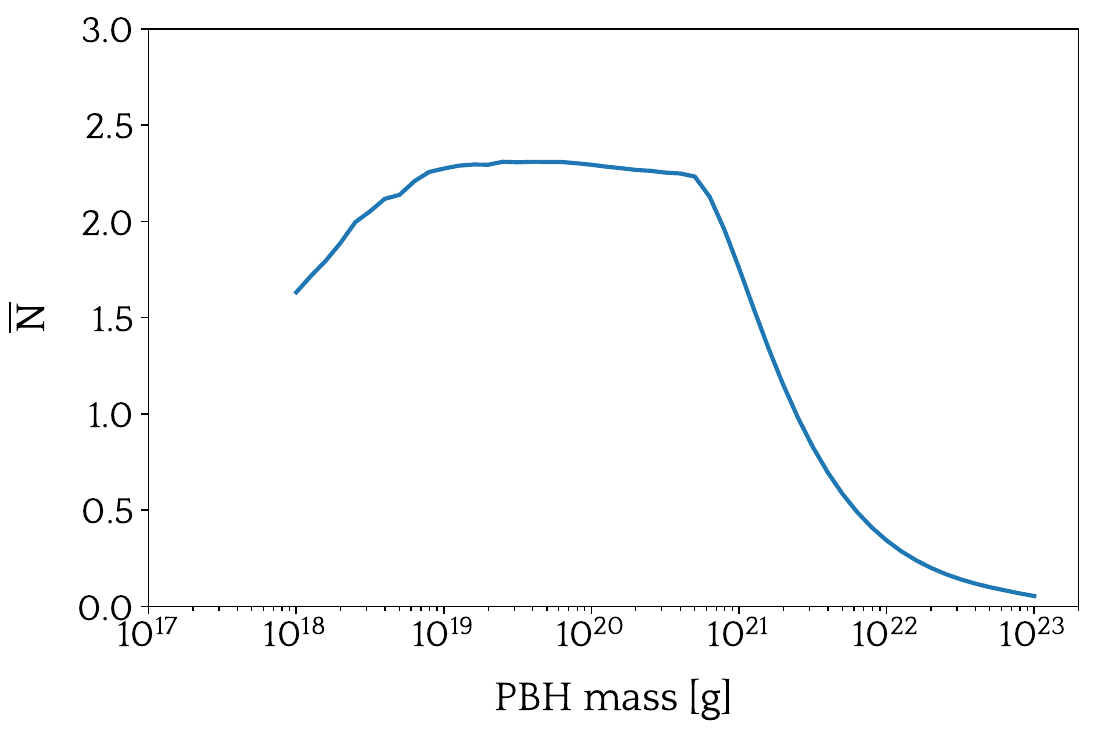}
  \caption{The mean number $\bar N$ of PBHs captured by a star of $1~M_\odot$
    at birth as a function of the PBH mass. The calculation assumes an
    isolated star at rest in an environment with the PBH mean mass density of
    $100$~GeV/cm$^3$ and velocity dispersion $7$~km/s. The plot is based on
    the same simulation as Fig.~1 of Ref.\cite{Esser:2022owk}.}
  \label{fig:captured_mass}
\end{figure}

The phase space region sampled by the initial conditions contains some total
number and total mass of PBH. Knowing the fraction of the
successful (that is, leading to capture) trajectories from simulations, one
may immediately determine the mean captured number $\bar N$ and the mean
captured mass $\bar M = m\bar N$ of PBHs. For a $1~M_\odot$ star the result is
shown in Fig.~\ref{fig:captured_mass}, which   
is the same as Fig.~1 of Ref.\cite{Esser:2022owk}, except $\bar N$ is
shown instead of $\bar M$.

On the horizontal part of the plot in the range roughly $10^{19}-10^{21}$~g,
the mean captured number is limited by the cooling time. There it is in
agreement with eq.~(\ref{eq:N-tcool}) which predicts no dependence on the PBH
mass. The drop at around $10^{21}$~g occurs where the requirement of not being
perturbed by the nearby stars becomes more important than that of the cooling
time. This part of the curve depends on the assumed ambient star density $n_*$;
the value $n_*=0.009$~pc$^{-3}$ as inferred from measured parameters of
the Triangulum II dwarf galaxy was used on this plot. 

At PBH masses smaller than $\sim 10^{19}$~g the mean captured number starts to
decrease. This can be traced back to breaking of the approximations used to
derive eq.~(\ref{eq:N-tcool}), in particular of the assumption that the
trajectories with $a\gg b$ give the dominant contribution.

\subsection{Mean captured number: compact stars}
\label{sec:mean-N-compact}

Compact stars form as a result of evolution of main sequence stars. Massive
stars with masses in the range of $8M_\odot \lesssim M_* \lesssim 15M_\odot$
during their main sequence phase typically end their evolution as neutron
stars, the lighter ones become white dwarfs, while the heavier ones collapse
directly into black holes. The lifetimes of massive stars decrease as the mass
grows, reaching about $\sim 10$~Myr at the heavy mass end.  If a main sequence
star at its birth has trapped one or more PBHs on bound star crossing orbits,
and if they get captured inside the progenitor star and sink to the center
before it evolves into a compact remnant, they may be inherited by the latter.

This two-step process has been considered in
Refs.\cite{Capela:2012jz,Capela:2014ita} where the amount of captured PBHs was
estimated for neutron stars and white dwarfs in the assumption that at the
first stage the PBHs do not destroy the progenitor. This assumption is not
crucial for deriving constraints based on the neutron star and white dwarf
disappearance because the destruction of a progenitor star has the same
effect.  Progenitors of various masses have been considered. The results were
presented in the form of expected constraints on the fraction of PBHs in the
total amount of dark matter that one could impose provided a population of
neutron stars and/or white dwarfs is found in dwarf galaxies. Unfortunately,
this is not presently the case. Note, however, that perturbations by nearby
stars were not taken into account in the analysis of
Refs.\cite{Capela:2012jz,Capela:2014ita}.

\section{Implications for constraints on PBH abundance}
\label{sec:expected-constraints}

As already mentioned, the destruction of stars --- main sequence or compact
--- by PBHs may potentially have several different signatures that may be used
to constrain (or detect) the PBH abundance. Of these, a mere star
disappearance due to their conversion into sub-solar mass black holes appears
the most robust one. We will now consider this signature in more detail; other
signatures will be briefly discussed in the concluding section. We will focus
on main sequence stars because they are much easier to observe, while the
deficit of compact stars due to their transmutation into black holes is much
harder to establish experimentally. In fact, as of today, no neutron stars or
white dwarfs have been detected in dwarf galaxies.

In the previous sections we have seen that PBHs of asteroid masses are very
difficult to capture inside stars. All the mechanisms that we have considered
require cosmologically long time and may lead to sizable probability of
capture only in exceptional environments. Both the probability of the direct
capture, eq.~(\ref{eq:direct-cap-NS}), and the capture at birth,
eq.~(\ref{eq:Ncap-num}), are proportional to $\rho_\text{DM}
/\sigma^3$. Following Ref.\cite{Esser:2022owk} one may classify the existing
environments according to their ``merit factor'' $\eta$ defined as
\[
\eta = {\rho_\text{DM}\over 100~\text{GeV/cm}^3}
\left(
{7~\text{km/s} \over \sqrt{2} \sigma}
\right)^3,
\]
where $\sqrt{2}$ accounts for the fact that when both PBHs and stars
have the same velocity dispersion $\sigma$, the relative velocity has the
dispersion $\sqrt{2}\sigma$. In the environments with $\eta=1$ 
eqs.(\ref{eq:direct-cap-NS}) and (\ref{eq:Ncap-num}) apply with
$\rho_\text{DM}$- and $\sigma$-dependent factors equal to 1. 

Of all the astrophysical environments with experimentally confirmed DM density
only some ultra-faint dwarf galaxies (UFDs) reach the values of $\eta$ close
to 1. Among the most promising UFDs are the Triangulum II ($\eta \geq 0.95$),
Tucana III ($\eta \geq 0.51$), Draco II ($\eta \geq 0.39$), Segue 1 ($\eta =
0.39$) and Grus I ($\eta = 0.37$). These numbers are calculated from actually
measured parameters of these galaxies \cite{Esser:2022owk}; the inequalities
are used where only the upper bounds on the velocity dispersion exist. For
comparison, the Milky Way in the solar neighborhood has $\eta \sim 2\times
10^{-8}$.

The PBH capture is a random uncorrelated process and therefore the probability
to capture a given number of PBHs follows the Poisson distribution with the
expectation equal to the mean captured number $\bar N$. For a solar mass star
this quantity was calculated numerically in Ref.~\cite{Esser:2022owk}; it is
shown in Fig.~\ref{fig:captured_mass} for the Trinagulum II galaxy which has
$\eta\simeq 1$.  In the mass range $\sim 10^{19}-10^{21}$~g the mean captured
number is $\bar N \simeq 2.2$ assuming PBHs constitute all of the DM. This
corresponds to the survival probability (the probability to capture zero PBHs)
$p_0 = \exp(- \bar N)\simeq 0.11$, meaning that the stars actually observed in
this galaxy would constitute only 11\% of their original number as formed
about 10~Gyr ago, while 89\% would have been destroyed by PBHs.

\begin{figure}[H]
  \centering
  \includegraphics[width=0.8\textwidth]{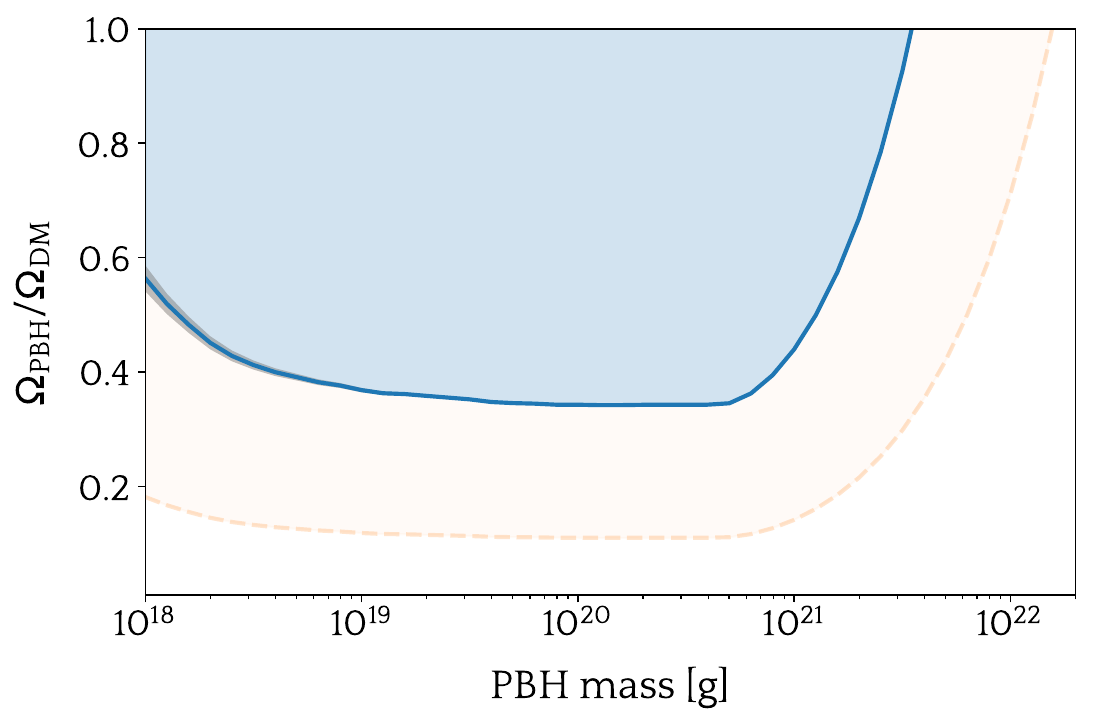}
  \caption{Upper limits on the PBH abundance
    $\Omega_{\rm PBH}/\Omega_{\rm DM}$ as a function of the PBH mass for the
    Triangulum II dwarf galaxy {\em assuming} that no more than 50\% (blue) or 20\%
    (orange) of stars were destroyed by PBHs. 
    Figure adopted from
    Ref.\cite{Esser:2022owk}. }
  \label{fig:constraints}
\end{figure}

If PBHs constitute only a fraction $f=
\Omega_\text{PBH}/\Omega_\text{DM}$ of the DM, the mean captured number
would be $f\bar N$, and the survival probability would be higher for smaller
$f$. Requiring that no more than a certain fraction of stars has been
destroyed translates into constraints on $f$ \cite{Esser:2022owk}. The result
is shown in Fig.~\ref{fig:constraints} for two assumed values of
the maximum destroyed fraction of stars equal to 50\% and 20\%.

\begin{figure}[h]
  \centering
  \includegraphics[width=0.8\textwidth]{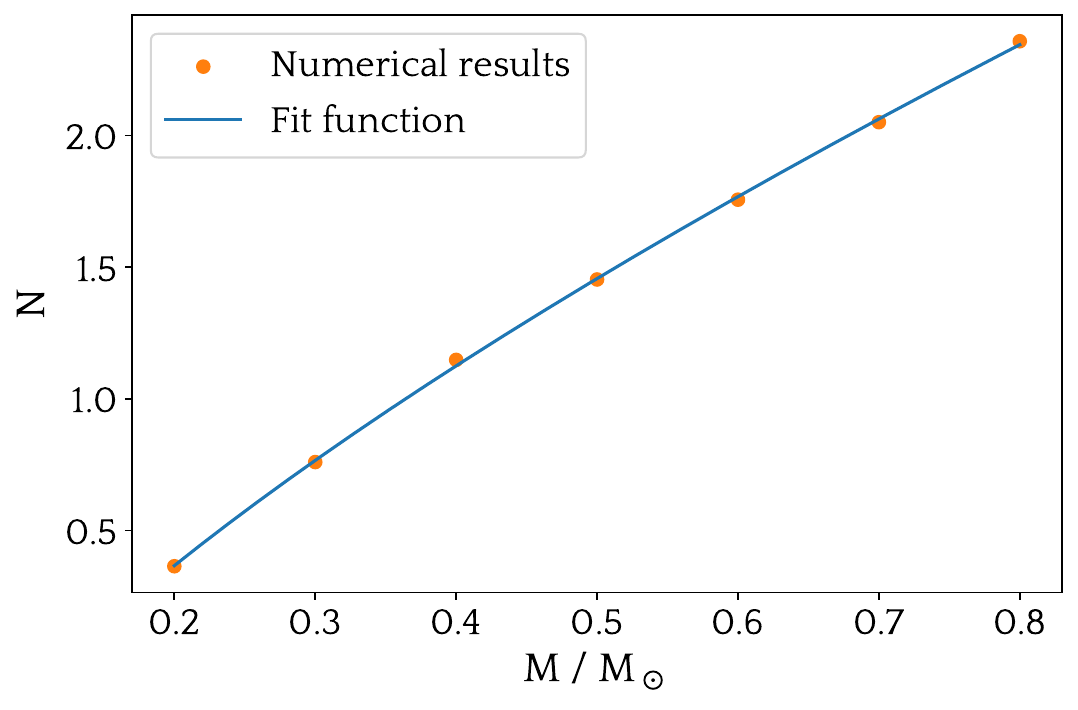}
  \caption{The mean captured number $\bar N$ as a function of the star mass $M$
    assuming $\eta=1$ and $f=1$. The dots show the numerical result, the blue
    line represents the fit by the function $a (M/M_\odot)^b+c $ with
    $a=3.76$, $b=0.685$ and $c=-0.883$. Figure adapted from
    Ref.\cite{Esser:2023yut}. }
  \label{fig:N-of-star-mass}
\end{figure}

It is not straightforward to infer the maximum allowed fraction of destroyed
stars from present-day observations as this would involve modeling of a dwarf
galaxy history. While such models exist, they have large uncertainties. A way
around has been proposed in Ref.\cite{Esser:2023yut}; it makes use of the fact
that the destruction probability grows with the mass of the star. This is
illustrated in Fig.~\ref{fig:N-of-star-mass} which shows the dependence of the
mean captured number of PBHs on the star mass for the case $\eta=f=1$. This
dependence is close to linear, and therefore the survival probability
decreases exponentially with the star mass. Thus, the presence of PBHs, apart
form reducing the number of stars, also strongly impacts the {\em stellar mass
  function} depleting its high-mass end. 

Known UFD galaxies (see Ref.~\cite{Simon} for a review) have from hundreds to
thousands of individual stars resolved with masses typically going down to
$0.5M_\odot$, and down to $0.2M_\odot$ in some cases. Moreover, it has been
shown \cite{jwst} that the James Webb Space Telescope can resolve stars down
to $0.09M_\odot$ in the dwarf galaxy Draco II at a distance of $\sim 20$~kpc.
From these data one can, in principle, reconstruct the stellar mass function
with a good accuracy and in a sufficiently wide  mass
range. Importantly, with the already existing technology, these data may be
extended to low masses $\lesssim 0.2 M_\odot$ where the effect of PBHs should
be negligible.

Modeling of the stellar mass distribution at birth, or initial mass
function, is a standard tool in the galactic astrophysics 
\cite{Kroupa,Chabrier}. The UFDs are of particular interest in this context
because, according to current understanding, their stars have all been
formed in a single burst some 12.8~Gyr ago. The long-lived ones --- those with
masses below $\sim 0.8M_\odot$ --- should therefore have preserved their
initial mass distribution until present.

The destruction of stars by PBHs modifies this picture. Importantly, the
effect of PBHs switches on and off very sharply (exponentially) as the merit
factor goes through the value $\eta\approx 1$. Thus, comparing several UFDs
with $\eta$ scattered around 1 should give a good handle on the parameters of
the unmodified stellar mass function, as well as on its possible
modifications due to the star destruction by PBHs. 

Preliminary estimates of the sensitivity of this approach have been made in
Ref.\cite{Esser:2023yut} using simulated data. A star population was generated
with the Kroupa mass function with the standard parameters and the number of
stars corresponding to a typical UFD, thus mimicking closely the observational
data. These mock data then have been fitted to a model based on a generic mass
function model (broken power law or log-normal) with a few free parameters
varying in the ranges inferred from observations. The model also included the
destruction of stars by PBHs controlled by one more free parameter, the
fraction of PBHs $f$ varying between 0 and 1. The Bayesian method was used to
infer the likely values of all the parameters, including $f$. The results show
that the method is sensitive enough to exclude large values of $f$ down to
$0.8$ at more than $3\sigma$ level. It appears therefore quite feasible to
exclude 100\% of the DM being composed of PBHs with already existing
observations.

\section{Summary and concluding remarks}
\label{sec:summary-conclusions}

Primordial black holes are an appealing candidate for the dark
matter. However, large parts of the potentially available mass range are
constrained by various arguments, and only the so-called asteroid mass window
from $\sim 10^{17}$g to $\sim 10^{23}$g remains fully allowed until now. We
have argued that capture of PBHs by main sequence stars in dwarf galaxies may
be a promising way to probe this mass range. Moreover, it is conceivable that
this can be done with the existing technologies, if not with already existing
data.

The approach that we have outlined consists, essentially, in comparing the
stellar mass functions of dwarf galaxies with low and high values of the merit
factor $\eta$. For that good quality data on stellar populations are
required. Such data already exist for some of the galaxies with low $\eta$
where stars have been resolved down to $0.2M_\odot$. For high $\eta$ galaxies,
notably for the ones listed in Sect.~\ref{sec:expected-constraints}, the
current observations go down to $0.5M_\odot$. This may or may not be
sufficient to firmly exclude the possibility of all of the DM consisting of
PBHs, $\Omega_\text{PBH}/\Omega_\text{DM} =1$. Better observations of these
galaxies extending to smaller star masses will be very important for this
purpose.

We have focused on the most straightforward signature of the star destruction
by PBHs, the mere star disappearance. But there are likely other
signatures. If a sizable fraction of stars in a dwarf galaxy is destroyed by
PBHs, this would create a population of sub-solar mass black holes. They are
not directly visible, but may perhaps be detected in gravitational waves.  It
is likely that the star destruction itself is accompanied by a supernova-type
energy release. If this is a short event that might have happened any
time in the last 10 Gyr, we have to be very lucky to observe and identify one
today, unless it has features that make these explosions unambiguously
distinguishable from other supernova events. Alternatively, last stages of
the star accretion onto a black hole may be relatively long and luminous, in
which case we have a good chance to detect and maybe identify such an event as
being due to a PHB capture.

\acknowledgement{The author is indebted to N.~Esser for the help with making
  Fig.~\ref{fig:captured_mass} and for comments on the manuscript. This work
  is supported in part by the Institut Interuniversitaire des Sciences
  Nucl\'eaires (IISN) Grant No. 4.4503.15. }

\bibliographystyle{unsrt}
\bibliography{authorsample.bib}

\end{document}